\documentclass[reprint,superscriptaddress,amsmath,amssymb,prb]{revtex4-1}
\usepackage[dvipdfmx]{graphicx}
\usepackage{dcolumn}
\usepackage{bm}
\usepackage{color}
\usepackage{amssymb}
\usepackage{comment}

\begin{document}

\title{Determining the optimum thickness for high harmonic generation from nanoscale thin films: an \textit{ab initio} computational study}

\author{Shunsuke Yamada}
\affiliation{Center for Computational Sciences, University of Tsukuba, Tsukuba 305-8577, Japan}
\author{Kazuhiro Yabana}
\affiliation{Center for Computational Sciences, University of Tsukuba, Tsukuba 305-8577, Japan}

\date{\today}

\begin{abstract}
We theoretically investigate high harmonic generation (HHG) from silicon thin films with thicknesses from 
a few atomic layers to a few hundreds of nanometers, to determine the most efficient thickness for producing 
intense HHG in the reflected and transmitted pulses.
For this purpose, we employ a few theoretical and computational methods. 
The most sophisticated method is the \textit{ab initio} time-dependent density functional theory coupled 
with the Maxwell equations in a common spatial resolution.
This enables us to explore such effects as the surface electronic structure and light propagation, as well as
electronic motion in the energy band in a unified manner.
We also utilize a multiscale method that is applicable to thicker films.
Two-dimensional approximation is introduced to obtain an intuitive understanding of the thickness 
dependence of HHG. 
From these \textit{ab initio} calculations, we find that the HHG signals are the strongest in films with thicknesses of 
2--15 nm, which is determined by the bulk conductivity of silicon. 
We also find that the HHG signals in the reflected and transmitted pulses are identical in such thin films.
In films whose thicknesses are comparable to the wavelength in the medium, the intensity of 
HHG signals in the reflected (transmitted) pulse is found to correlate with the magnitude of the electric field at the 
front (back) surface of the thin film.
\end{abstract}
\maketitle

\section{Introduction}

Following progress in high harmonic generation (HHG) in atoms and molecules
that enabled the production of attosecond pulses \cite{Corkum1993, Lewenstein1994, Brabec2000, Krausz2009}, 
HHG in bulk crystals has attracted great interest during the last decade  \cite{Ghimire2019, Li2020}.
Experimental studies have achieved HHG in thin films of various materials and laser pulses
\cite{Ghimire2011, Schubert2014, Vampa2015, Hohenleutner2015, Luu2015, Langer2017, You2017, Shirai2018, Vampa2018, Orenstein2019}.
These are extended to monatomic two-dimensional (2D) layers, such as graphene
\cite{Yoshikawa2017, Yoshikawa2019},
and to metasurfaces, periodic 2D structures composed of nanoscale objects \cite{Liu2018}.
HHG from bulk crystals is expected to be very intense compared with 
those from atoms in the gas phase because of the high atomic density.
This indicates that the HHG from solids is favorable in applications to develop,
for example, compact devices of XUV light sources.

Theoretical studies in the past mostly focused on electronic motion
induced by a pulsed electric field prepared in advance \cite{Yu2019}.
For sufficiently thick films, it is legitimate to consider electronic motion
in an infinitely periodic crystalline system in three dimensions induced by a 
spatially uniform electric field.
Calculations employing theories of varying complexity, such as one-dimensional model,
\cite{Hansen2017, Hansen2018, Jin2019, Ikemachi2017},
time-dependent Schr\"odinger equation \cite{Wu2015, Apostolova2018},
density matrix models \cite{Ghimire2012, Vampa2014, Vampa2015-2},
Floquet theory \cite{Faisal1997, Faisal2005}
and \textit{ab initio} descriptions such as the
time-dependent density functional theory (TDDFT) 
\cite{Otobe2012, Otobe2016, Tancogne-Dejean2017, Floss2018}, have been developed.
Theories to describe HHG from monatomic layers and bulk surfaces have also been
developed \cite{Georges2005, LeBreton2018}.
Using these theories, electronic motion in the wavenumber ($k$) space 
based on the energy band picture has been investigated and classified as  inter- and intra-band motions \cite{Vampa2014}. 
It has been clarified that there are rich manifestations of the crystalline structure of HHG 
through the band structure, selection rules, and dependence on the polarization direction
\cite{Langer2017, You2017, Neufeld2019}.

To investigate HHG in bulk materials, the propagation effect is also important. 
The light-propagation effect on HHG has been considered already in the case of atomic gases \cite{Popmintchev2009}. 
However, the effect should be significant primarily in bulk solids, owing to their high atomic density.
The propagation effect in HHG from thin films has been investigated recently.
Experimentally, HHG in reflected and transmitted pulses has been measured and compared 
for films of several different thicknesses \cite{Vampa2018-2, Xia2018}.
Theoretically, multiscale calculations coupling a coarse-graining electronic motion
with light propagation have been conducted \cite{Ghimire2012, Zhokhov2017, Floss2018, Orlando2019, Kilen2020}. 
It has been shown that the light propagation
works to produce a clear HHG spectrum, even when the spectrum is unclear in the unit-cell calculation \cite{Floss2018}. It has also been shown that the HHG in the transmitted pulse decreases 
by several orders of magnitude when emitted from thin films of micro-meter thickness \cite{Kilen2020}.  

Although there has been progress on the propagation effect, as described above, 
we consider that even the following basic questions have not yet been answered:
(1) At what film thickness is HHG emitted with the maximum intensity?
(2) What are the differences and similarities between HHGs in reflected and transmitted pulses?
The principal purpose of this paper is to answer these questions unambiguously.

We theoretically describe HHG in Si thin films of various thicknesses,
from a few atomic layers to a few hundreds of nanometers, that are comparable
to the wavelength of the incident pulse in the medium.
We utilize a few theoretical and computational frameworks based on
\textit{ab initio} TDDFT \cite{Runge1984, Ullrich2012} coupled with
 the Maxwell equations. In the most sophisticated method, we simultaneously solve the 
time-dependent Kohn-Sham (TDKS) equation for the electronic 
motion in TDDFT and the Maxwell equations for electromagnetic fields 
with a common spatial resolution. We call this the single-scale Maxwell-TDDFT method \cite{Yamada2018}.
The method is applicable to films with thicknesses of less than a few tens of nanometers
because of its high computational cost.
For thicker films, we utilize a multiscale Maxwell-TDDFT method, in which
a microscopic TDKS equation is coupled with the macroscopic Maxwell equations 
with a coarse-graining approximation \cite{Yabana2012}. 
A 2D approximation \cite{Yamada2018} is also introduced 
to obtain an intuitive understanding of the thickness dependence of HHG.
Calculations using these methods provide a unified understanding on the HHG 
in nanoscale thin films.

This paper is organized as follows: 
Sec.~\ref{sec:method} describes the theoretical and computational methods
based on TDDFT.
In Sec.~\ref{sec:results}, the calculation results are presented. The thickness dependence of
HHG, as well as the relation between HHGs in  reflected and transmitted pulses, are discussed.
Finally, a conclusion is presented in Sec.~\ref{sec:conclusion}.

\section{Theoretical methods \label{sec:method}}

\subsection{Problem setup and method summary}

We consider  irradiation of a free-standing thin film of thickness $d$ in a vacuum 
by an ultrashort light pulse of a linearly polarized plane wave at normal incidence.
As a prototypical material, we selected Si thin films of various thickness, 
from a few atomic layers to a few hundreds of nanometers.
We assume that the thin film is infinitely periodic and macroscopically isotropic in 2D.
The  propagation and polarization directions of the incident light are taken  to be along the $z$ and $x$ axes, respectively.
We assume that the reflected and transmitted pulses contain only a component in the $x$ direction, owing to  symmetry.

We express the asymptotic form of the vector potential in the following way using
the $x$ component of the vector potential $A(z,t)$:
\begin{equation}
    A(z,t)=
\begin{cases}
  A^{\rm (i)}( t-z/c )
  + A^{\rm (r)} ( t+z/c ) & (z \rightarrow -\infty ),
\\
  A^{\rm (t)}( t-z/c ) & (z \rightarrow + \infty),
\end{cases}
\label{A_asymptotic}
\end{equation}
where $A^{\rm (i)}(t)$, $A^{\rm (r)}(t)$, and $A^{\rm (t)}(t)$  are the incident, reflected, and transmitted fields, respectively.
The $x$ component of the electric field, $E(t)$, is related to the vector potential by $E(t)=-(1/c) d A(t)/dt$.
The reflected and transmitted electric fields, $E^{\rm (r)}(t)$ and  $E^{\rm (t)}(t)$, contain the high-order harmonics 
generated by the nonlinear light-matter interaction in the film.

To calculate HHG spectrum from a thin film theoretically, 
it has been commonly considered electronic motion in a unit cell of crystalline solids 
induced by an applied electric field whose time profile is prepared in advance.
There are, however, a number of effects that should be considered further.
We investigate the following three effects in this paper.
First is an effect of the surface electronic structure, that is, a change of 
the electron band structure at the surfaces from that of the bulk system.
The other two effects are related to the difference between the incident electric field
$E^{\rm (i)}(t)$ and the electric field that actually acts on electrons in the thin film.
We consider two effects that cause this difference.
One is the effect of the light propagation.
A strong incident laser pulse is modulated during the propagation due to nonlinear light-matter interactions, and
HHG pulses that are generated inside the thin film also suffer from strong modulation and absorption 
during propagation before they exit the medium at the surfaces.
The other is the effect of the conductivity of the thin film.
Even when the thin film is so thin that the electric field inside the thin film can be regarded as uniform at the macroscopic scale, 
the electric field inside the film is different from that of the incident pulse due to the conductivity.
Specifically, the current that flows in the film causes reflection from the film, and the electric field inside the medium 
is equal to the sum of the incident and  reflected electric fields, not the incident field alone.

For a theoretical and computational description of HHG from a thin film,
we  utilize four \textit{ab initio} approaches based on TDDFT \cite{Runge1984, Ullrich2012}, which are summarized in Table I and explained in the
following subsections.
The simplest approach is the single unit-cell method, in which we solve the TDKS equation for electronic motion
in a unit cell of the crystalline solid. 
In this description,  the electric field of the incident pulse is used as an applied field.
The most extensive and sophisticated approach is the single-scale Maxwell-TDDFT method \cite{Yamada2018},
in which we solve the Maxwell equations for the electromagnetic fields and the TDKS equation 
for the electronic motion simultaneously using a common spatial grid. 
Here, all three effects mentioned above are included.
In the multiscale Maxwell-TDDFT method \cite{Yabana2012}, macroscopic light propagation and microscopic electronic
motion are coupled using a coarse-graining approximation.
In the practical calculation, the Maxwell equations and TDKS equation are solved simultaneously 
using different spatial grids. While the effect of the surface electronic structure is not included, 
this method can describe HHG from thick films that cannot be accessed by the single-scale Maxwell-TDDFT method.
Finally, the 2D approximation \cite{Yamada2018} is introduced as an approximation to the multiscale Maxwell-TDDFT method,
treating the thin film as an infinitely thin in macroscopic scale.
This method will be useful to understand the thickness dependence of the HHG.

\begin{table*}
\caption{\label{tab:method} 
Methods that will be used to describe HHG and their effectiveness to take account of various effects.
}
\begin{ruledtabular}
  \begin{tabular}{lcccc}
    Theory & Single-scale Maxwell-TDDFT & Multiscale Maxwell-TDDFT & 2D approximation & Single unit-cell \\
    Treatment of EM fields & Microscopic  & Macroscopic & 2D macroscopic & None \\
      \hline
     Bulk electronic structure & Yes & Yes & Yes & Yes \\
    Surface electronic structure & Yes & No & No & No \\
    Light propagation & Yes & Yes & No & No \\
    Conduction & Yes & Yes & Yes & No
  \end{tabular}
\end{ruledtabular}
\end{table*}

\subsection{Single unit-cell method}

First, we  consider a method describing the electronic motion in a unit cell of a crystalline solid
under a spatially uniform electric field \cite{Bertsch2000, Otobe2008}. This method describes electronic motion based on the 
energy-band picture, and has been utilized extensively  to describe phenomena related to
nonlinear and ultrafast dynamics of electrons such as nonlinear susceptibilities \cite{Uemoto2019},
nonlinear energy transfer \cite{Yamada2019}, attosecond dynamics \cite{Wachter2014, Schultze2014, Yamada2020}, and HHG from various solids 
\cite{Otobe2012, Otobe2016, Tancogne-Dejean2017, Floss2018}.
Within the dipole approximation, the electron dynamics are
described using the  Bloch orbital of the bulk system, $u^{\rm bulk}_{n{\bm k}}({\bm r},t)$,
where the wave vector ${\bm k}$ runs over the three-dimensional (3D) Brillouin zone of the unit cell.
The TDKS equation for the Bloch orbitals $u^{\rm bulk}_{n{\bm k}}({\bm r},t)$ is written as
\begin{eqnarray}
    i\hbar \frac{\partial u^{\rm bulk}_{n{\bm k}}({\bm r},t)}{\partial t}
    = \biggl\{ \frac{1}{2m} \left( -i\hbar \nabla + \hbar{\bm k} + \frac{e}{c} \hat{\bm x}A^{\rm (i)}(t) \right)^2 
      \nonumber \\ 
      - e\phi({\bm r},t) + \delta \hat V_{\rm{ion}} + V_{\rm{xc}}({\bm r},t) \biggr\} u^{\rm bulk}_{n{\bm k}}({\bm r},t).\quad
      \label{eq:tdks}
\end{eqnarray}
Here, we use the vector potential of the incident pulse, $A^{\rm (i)}(t)$, as the applied electric field.
The electric scalar potential $\phi({\bm r},t)$ includes the Hartree potential from the electrons and
the local part of the ionic  potential.
$\delta \hat V_{\rm{ion}}$ and $V_{\rm{xc}}({\bm r},t)$ are the nonlocal part of the ionic pseudopotential \cite{Troullier1991}
and the exchange-correlation potential \cite{Perdew1981}, respectively. 
All of the calculations within this paper  are performed within the adiabatic local-density approximation\cite{Onida2002}.
We ignore the exchange-correlation effect for the vector potential, $A_{\rm xc}(t)$, for simplicity \cite{Ullrich2012}.

The electric current density averaged over the unit cell is calculated as follows:
\begin{eqnarray}
&& {\bm J}(t) = -  \frac{e}{m} \int_{\Omega} \frac{d \bm r}{\Omega} \sum^{\rm occ}_{n{\bm k}} u^{\rm bulk \ast}_{n{\bm k}} ({\bm r},t) 
\nonumber \\ 
&& \times \left( 
-i\hbar \nabla + \hbar{\bm k} + \frac{e}{c} \hat{\bm x}A^{\rm (i)}(t)
\right) u^{\rm bulk}_{n{\bm k}} ({\bm r},t) + \delta {\bm J}(t),
\label{eq:Jav}
\end{eqnarray}
where $\Omega$ is the volume of the unit cell and the index $n$ runs over the occupied band in the ground state.
$\delta {\bm J}(t)$ is the current density  from the nonlocal part of the pseudopotential \cite{Bertsch2000}.
\begin{eqnarray}
\delta {\bm J}(t) &=& -\frac{e}{m} \int \frac{d{\bm r}}{\Omega} \sum_{n {\bm k}}^{\rm occ}
u^{\rm bulk \ast}_{n{\bm k}}({\bm r},t) e^{-i({\bm k} + (e/c)\hat {\bm x}A^{\rm (i)}(t)) {\bm r}} \nonumber\\
&& \times
\frac{[ {\bm r}, \delta \hat V_{\rm ion}]}{i\hbar} e^{i({\bm k} + (e/c) \hat {\bm x}A^{\rm (i)}(t)) {\bm r}}u^{\rm bulk}_{n {\bm k}}({\bm r},t).
\end{eqnarray}
In practice, the current contains only the $x$ component, parallel to the polarization direction of the incident pulse.
We later discuss how to relate this current density with the HHG spectrum from a thin film of thickness $d$.

\subsection{Single-scale Maxwell-TDDFT method}

We next consider two theoretical methods that are capable of describing the electronic motion and light propagation simultaneously.
The first is the single-scale Maxwell-TDDFT method, in which the electromagnetic fields are treated microscopically.
In the next subsection, we will present the second method, the multiscale Maxwell-TDDFT method,  in which the electromagnetic
fields are treated macroscopically. 

The single-scale Maxwell-TDDFT method\cite{Yamada2018} can fully  account for the effects of the light propagation,
 conduction, and  surface electronic structure.
In this method, the Maxwell equations for electromagnetic fields and the TDKS equation for electronic motion are solved 
simultaneously in the time domain using a common spatial grid.
In the present case, we consider an atomic configuration that is infinitely periodic in the $xy$ plane and  isolated
in the $z$ direction.
The electronic motion is described using the Bloch orbital in the slab approximation, $u^{\rm slab}_{n{\bm k}}({\bm r},t)$, where ${\bm k}$ is the 2D crystal wave vector.
This satisfies the following TDKS equation:
\begin{eqnarray}
    i\hbar \frac{\partial u^{\rm slab}_{n{\bm k}}({\bm r},t)}{\partial t}
    = \biggl\{ \frac{1}{2m} \left( -i\hbar \nabla + \hbar{\bm k} + \frac{e}{c} {\bm A}({\bm r},t) \right)^2 
      \nonumber \\ 
      - e\phi({\bm r},t) + \delta V_{\rm{ion}}({\bm r}) + V_{\rm{xc}}({\bm r},t) \biggr\} u^{\rm slab}_{n{\bm k}}({\bm r},t). \quad
\end{eqnarray}
The vector potential ${\bm A}({\bm r},t)$ and the scalar potential $\phi({\bm r},t)$ are 2D periodic and satisfy the Maxwell equations in the Coulomb gauge:
\begin{equation}
    \left( \frac{1}{c^2} \frac{\partial^2}{\partial t^2} - \nabla^2 \right) {\bm A}({\bm r},t)
    + \frac{1}{c} \frac{\partial}{\partial t} \nabla \phi({\bm r},t) =  \frac{4\pi}{c} {\bm j}({\bm r},t),
    \label{eq:maxwell_micro}
\end{equation}
\begin{equation}
    \nabla^2 \phi({\bm r},t) = - 4\pi \rho ({\bm r},t).
\end{equation}
The vector potential ${\bm A}({\bm r},t)$ is smoothly connected to the asymptotic field of Eq.~(\ref{A_asymptotic})
at planes that are sufficiently separated from the medium. The scalar potential is chosen to be periodic in the $z$ direction
as well as in the $x$ and $y$ directions.
As mentioned above, all the dynamic variables, Bloch orbitals, and scalar and vector potentials are calculated 
using a common spacial grid without coarse graining. In this sense, both the vector and scalar potentials
are treated microscopically in this approach.

The charge density $\rho({\bm r},t)$ and the electric current density  ${\bm j}({\bm r},t)$ are derived from the Bloch orbitals, as follows:
\begin{eqnarray}
    \rho({\bm r},t) &=& \rho_{\rm ion}({\bm r}) -e \sum_{n{\bm k}}^{\rm{occ}} |u^{\rm slab}_{n{\bm k}}({\bm r},t)|^2 ,
    \label{eq:density} \\
    {\bm j}({\bm r},t) &=& -\frac{e}{m} {\rm Re} \sum_{n{\bm k}}^{\rm occ}  u_{n{\bm k}}^{\rm slab \ast}({\bm r},t)
    \Bigl( -i\hbar \nabla \nonumber \\
    && + \hbar{\bm k} + \frac{e}{c} {\bm A}({\bm r},t) \Bigr) u^{\rm slab}_{n{\bm k}}({\bm r},t),
    \label{eq:current}
\end{eqnarray}
where $\rho_{\rm ion}({\bm r})$ is the  charge density of the ion cores.
In this method, we ignore the current from the nonlocal part of the pseudopotential\cite{Yamada2018}.

Because the Bloch orbitals $\{u^{\rm slab}_{n{\bm k}}({\bm r},t)\}$ are defined in the calculation box that includes entire film as well
as the vacuum region, the size of the computation rapidly grows as the thickness of the film increases. 
In terms of the computational cost, applications of the single-scale Maxwell-TDDFT method is limited to thin films of
thickness less than several tens of nanometers.

\subsection{Multiscale Maxwell-TDDFT method}


In the multiscale Maxwell-TDDFT method \cite{Yabana2012}, macroscopic light propagation and microscopic 
electronic motion are coupled using a coarse-graining approximation.
The light propagation is described using the following macroscopic wave equation:
\begin{equation}
 \left (\frac{1}{c^2} \frac{\partial^2}{\partial t^2}  - \frac{\partial^2}{\partial Z^2} \right) A_{Z}(t)
= \frac{4\pi}{c}   J_{Z}(t),
\label{eq:multiscale}
\end{equation}
where $Z$ is the macroscopic coordinate. This wave equation is solved using a one-dimensional grid.
At each grid point of $Z$, a bulk system with 3D periodicity is considered. 
The electronic motion at each grid point of $Z$ is described using 3D-periodic Bloch orbitals, $u^{\rm bulk}_{n{\bm k},Z}({\bm r},t)$,
which satisfy the TDKS equation:
\begin{eqnarray}
    i\hbar \frac{\partial u^{\rm bulk}_{n{\bm k},Z}({\bm r},t)}{\partial t}
    = \biggl\{ \frac{1}{2m} \left( -i\hbar \nabla + \hbar{\bm k} + \frac{e}{c} \hat{\bm x}A_{Z}(t) \right)^2 
      \nonumber \\ 
      - e\phi_Z({\bm r},t) + \delta V_{\rm{ion}}({\bm r}) + V_{\rm{xc},Z}({\bm r},t) \biggr\} u^{\rm bulk}_{n{\bm k},Z}({\bm r},t).\quad
\label{eq:msKS}
\end{eqnarray}
From the Bloch orbitals, the averaged electric current density $J_Z(t)$ is calculated in the same manner as Eq.~(\ref{eq:Jav}), but replacing $A^{\rm (i)}(t)$ with $A_{Z}(t)$.

At the beginning of the calculation, the Bloch orbitals at each grid point $Z$, $u^{\rm bulk}_{n {\bf k},Z}({\bm r},t)$, are set to the ground state.
The vector potential $A_Z(t)$ is set to include only the incident pulse in the vacuum region.
By solving Eqs.~(\ref{eq:multiscale}) and (\ref{eq:msKS}) simultaneously, we can evolve both $A_Z(t)$ and
$u^{\rm bulk}_{n{\bm k},Z}({\bm r},t)$ simultaneously.
We note that microscopic electronic systems at different $Z$ positions interact only through the vector potential $A_Z(t)$.
When the light pulse is sufficiently weak and thus the perturbation approximation is applicable,
the multiscale Maxwell-TDDFT method results in the ordinary macroscopic electromagnetism with
the constitutive relation given by the dielectric function in the linear response TDDFT \cite{Yabana2012}.
The multiscale Maxwell-TDDFT method has been successfully applied to investigate a number of extremely nonlinear 
and ultrafast phenomena including attosecond science \cite{Lucchini2016, Sommer2016}, 
saturable absorption \cite{Uemoto2021}, coherent phonon generation and detection \cite{Yamada2019-2, Yamada2020-2}, 
and initial stage of nonthermal laser processing \cite{Sato2015}.

In the multiscale Maxwell-TDDFT method, the effect of the light propagation can be taken into account
while the effect of the surface electronic structure cannot.
In compensation for the surface effect, the multiscale Maxwell-TDDFT method can be applied to thick films for
which the single-scale Maxwell-TDDFT method is not capable.
As we will show later, results of the multiscale Maxwell-TDDFT method coincide reasonably with those of the
single-scale Maxwell-TDDFT method when the surface effect is not significant.

\subsection{2D approximation}

For sufficiently thin films, we may assume that the macroscopic electric field is spatially uniform inside the thin film.
We call this the 2D approximation.
This approximation is useful to identify and distinguish the conductive effect from the propagation effect, 
and to understand the thickness dependence of the HHG.
The 2D approximation can be derived from the multiscale Maxwell-TDDFT method, as described below.
Alternatively, it can also be derived from the single-scale Maxwell-TDDFT method, as explained in Ref~\onlinecite{Yamada2018}.

We consider a film that is sufficiently thick to justify neglecting the effect of the surface electronic structure
and  sufficiently thin to regard the macroscopic electric field as spatially uniform in the film.
In the next section, we  see that there are thickness regions in which both assumptions are fulfilled simultaneously.
Under these assumptions, the current density in Eq.~(\ref{eq:multiscale}) may be treated as
\begin{equation}
J_Z(t) \simeq \delta(Z) J(t) d,
\label{eq:2Dcurrent}
\end{equation}
where $d$ is the thickness of the film and $J(t)$ is the current density averaged over the unit cell.
By inserting this current density into Eq.~(\ref{eq:multiscale}), we have
\begin{equation}
 \left (\frac{1}{c^2} \frac{\partial^2}{\partial t^2}  - \frac{\partial^2}{\partial Z^2} \right) A_Z(t)
= \frac{4\pi}{c} \delta(Z)  J(t)d.
\label{eq:2Deq}
\end{equation}
This equation\cite{Mishchenko2009} should be solved with the asymptotic form of the vector potential of Eq.~(\ref{A_asymptotic}).
In practice, the vector potential of Eq.~(\ref{A_asymptotic}) is the solution of Eq.~(\ref{eq:2Deq}),
except at $Z=0$. The reflected and  transmitted fields are determined by the connection conditions at $Z=0$.
From the continuity of the vector potential at $Z=0$, we have
\begin{equation}
A^{\rm (i)}(t) + A^{\rm (r)}(t) = A^{\rm (t)}(t).
\label{eq:2Dreftrans}
\end{equation}
By integrating  Eq.~(\ref{eq:2Deq}) once over $Z$, we obtain
\begin{equation}
\frac{dA^{\rm (t)}}{dt} =  \frac{dA^{\rm (i)}}{dt} + 2\pi d  J[A^{\rm (t)}](t)
\label{eq:2Dapprox}
\end{equation}
at $Z=0$, 
where we denote the current density as $J[A^{\rm (t)}](t)$ to indicate that this is the current density caused by the
vector potential $A_{Z=0}(t)=A^{\rm (t)}(t)$.
This is the basic equation of the 2D approximation.

We note that this 2D approximation is equivalent to a specific case of the multiscale Maxwell-TDDFT method.
If we take only a single grid point in the macroscopic coordinate $Z$ for the medium and choose the grid
spacing to be equal to $d$, the calculation using the multiscale Maxwell-TDDFT method coincides with the above 2D approximation.

From the 2D approximation, we obtain several useful understandings regarding the HHG from
very thin films.
The continuity equation (\ref{eq:2Dreftrans}) indicates that the HHG spectra of the reflected and transmitted pulses are equal.
Equation~(\ref{eq:2Dapprox}) indicates that the current density that creates HHG is produced by the transmitted pulse
$A^{\rm (t)}(t)$, not the incident pulse of $A^{\rm (i)}(t)$, and the transmitted pulse itself is modulated by the current density.
Even in very thin films such as monatomic layered films, the transmitted (and simultaneously, the reflected) fields are
modulated by the current that flows in the film. This modulation significantly affects the HHG spectrum, as shown in the next section.
We call this effect the conductive effect, as it is related to the electric current flowing in the thin film.

\subsection{High harmonic generation spectrum}

In the single-scale and multiscale Maxwell-TDDFT methods, we directly obtain
the vector potentials of the reflected and the transmitted pulses.
In the 2D approximation, we obtain the vector potentials of the transmitted pulse by Eq.~(\ref{eq:2Dapprox}).
The vector potential of the reflected pulse is obtained by using the relation Eq.~(\ref{eq:2Dreftrans}).
From the vector potentials, we obtain the respective electric fields by taking the time derivative.

For the single unit-cell method, we will use the following expression for the reflected and transmitted electric fields.
\begin{equation}
E^{\rm (r)}(t) = E^{\rm (t)}(t) - E^{\rm (i)}(t) = -\frac{2\pi d}{c} J[A^{\rm (i)}](t).
\end{equation}
This relation is derived if we replace $A^{\rm (t)}$ with $A^{\rm (i)}$ in $J[A^{\rm (t)}](t)$ of Eq.~(\ref{eq:2Dapprox}).

We evaluate the HHG spectra via the square of the Fourier-transformed electric field $|E^{\rm (r,t)}(\omega)|^2$, 
where $E^{\rm (r,t)}(\omega)$ is defined by,
\begin{equation}
   E^{\rm (r,t)}(\omega)= \int_{t_0}^{t_0 + T} dt \, e^{i\omega t} E^{\rm (r,t)}(t) \, f\left(\frac{t-t_0}{T}\right).
\end{equation}
Here, $f(x)\equiv 1-3 x^2 +2 x^3$ is a smoothing function\cite{Yabana2006} and $t_0$ is the initial time.
For the reflection pulse, $t_0$ is set to the initial time of the incident pulse.
For the transmitted pulse, the time delay by transmission is added to $t_0$ for thick films.
When we discuss the thickness dependence of HHG, we introduce the strength of the $n$th-order harmonics, as follows:
\begin{equation}
    I_n^{\rm (r,t)} = \int^{\left(n+\frac{1}{2}\right)\omega_0}_{\left(n-\frac{1}{2}\right)\omega_0} d\omega \,
    |E^{\rm (r,t)}(\omega)|^2 ,
    \label{eq:harmonics}
\end{equation}
where $\omega_0$ is the fundamental frequency.

\subsection{Numerical detail}

For numerical calculations, 
we utilize the open-source software SALMON (Scalable Ab initio Light-Matter simulator for
Optics and Nanoscience)\cite{Noda2019,SALMON_web} developed by our group. 
In this code, electronic orbitals as well as electromagnetic fields are expressed using a uniform 3D spatial grid.
The time evolution of the electron orbitals is carried out using the Taylor expansion method \cite{Yabana1996}.

In the single unit-cell method, we use a cubic unit cell that contains eight Si atoms with the side length of $a=0.543$ nm.
To express the Bloch orbitals, a spatial grid of $16^3$ points is used.
The 3D Brillouin zone is sampled by a $24^3$ $k$-point grid.
The time step is set to $2.5 \times 10^{-3}$ fs.

The same cubic unit cell is also used  in the 2D approximation and  multiscale Maxwell-TDDFT method.
In the multiscale Maxwell-TDDFT method, a uniform 1D grid is also introduced to describe the wave equation of Eq.~(\ref{eq:multiscale}), 
with the grid spacing  less than or equal to $6.25$ nm.

In the single-scale Maxwell-TDDFT method, the 2D-periodic box of the size of $a\times a\times(d + 4a)$ is used
where the size along the $z$ axis contains the vacuum region of $4a$ in the slab approximation.
The atomic positions of Si atoms are set at those positions in the bulk crystalline system, and 
the dangling bonds at the surfaces are terminated by hydrogen atoms.
A uniform 3D spatial grid is used with the same grid spacing as that in the cubic unit cell.
The 2D Brillouin zone of the single-scale Maxwell-TDDFT method is sampled by a $24 \times 24$ $k$-point grid.


\section{Results and discussion \label{sec:results}}

\subsection{HHG in reflected and transmitted pulses}

We employ the following time profile for the incident pulse:
\begin{eqnarray}
A^{\rm (i)}(t) = -\frac{c E_0}{\omega_0}  \, \sin \omega_0 t  \,\, \cos^6  \left( \frac{\pi t}{T} \right) ,\, (-\frac{T}{2}<t<\frac{T}{2}), \nonumber \\
\label{eq:incident}
\end{eqnarray}
where $E_0$ is the maximum amplitude of the electric field, $\omega_0$ is the average frequency, and $T$ is the pulse duration.
In the following calculations, we set $\hbar \omega_0=$ 1.5 eV and $T=$ 30 fs.

\begin{figure}
    \includegraphics[keepaspectratio,width=\columnwidth]{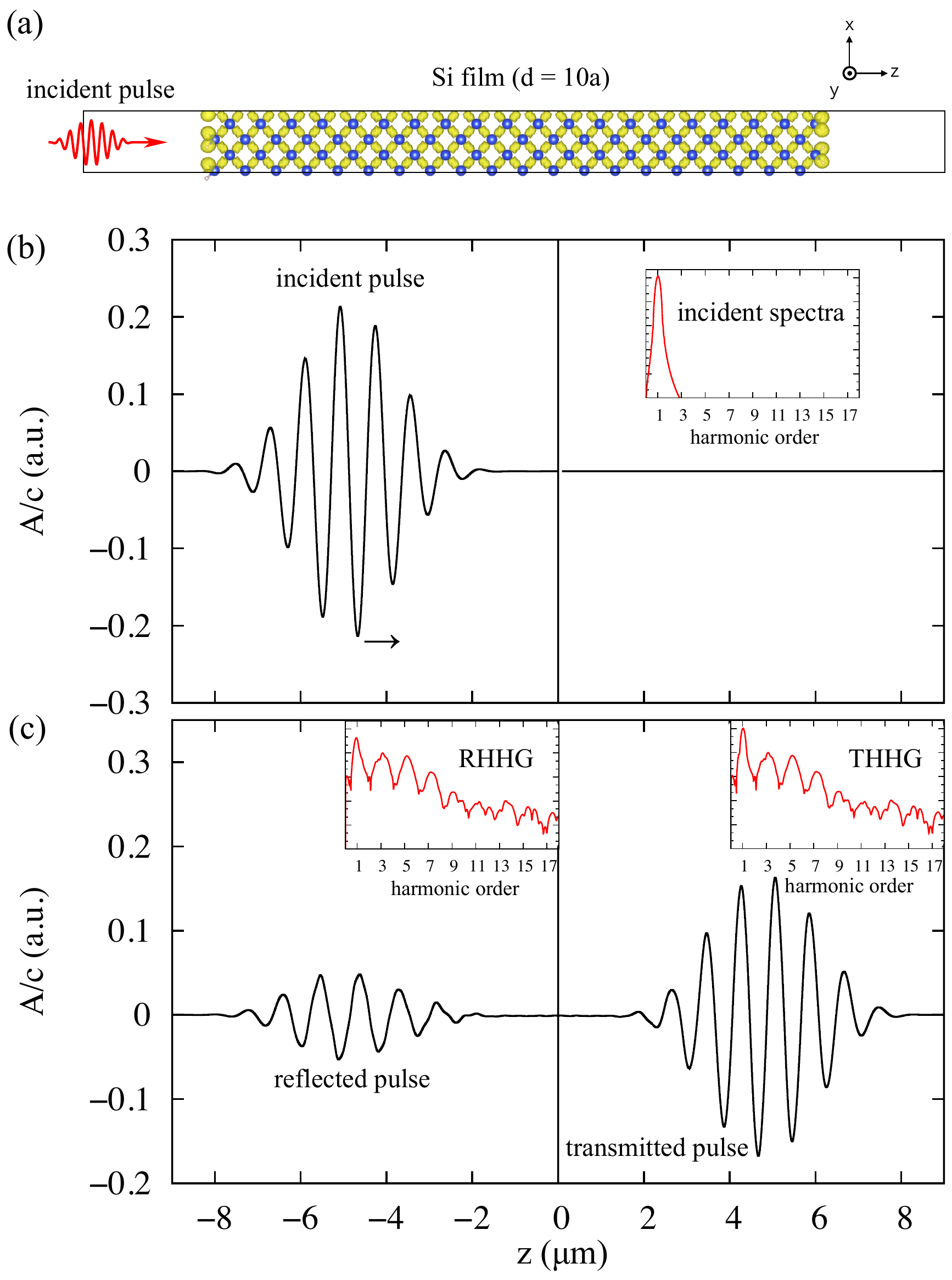}
    \caption{\label{fig:pulse} 
    Typical calculation using the single-scale Maxwell-TDDFT method for a Si thin film of thickness $d=10a$ (5.43 nm).
    (a) Electron density in the ground state in a plane containing Si atoms.
    (b) Initial pulse set to the left of the thin film located at $z=0$. The peak intensity of the pulse is
    set at $I=5\times 10^{12}$ W/cm${}^2$ at $t=0$.
    (c) Reflected and transmitted pulses at  $t=31.25$ fs. The insets show the corresponding HHG spectra.   
    }
\end{figure}

For a typical case, Figure~\ref{fig:pulse} shows snapshots of the incident pulse [Eq.~(\ref{eq:incident})] and 
the transmitted and the reflected pulses in the calculation using the single-scale Maxwell-TDDFT method 
for a film of thickness $d=10a$ (5.43 nm).
The maximum amplitude $E_0$ of the incident pulse is set to provide the peak intensity of $I=5\times 10^{12}$ W/cm${}^2$.
In Fig.~\ref{fig:pulse}(a), the electronic density at the ground state in a plane that includes Si atoms is shown.
In Fig.~\ref{fig:pulse}(b), the incident pulse is prepared to the left of the thin film located at $z=0$.
Figure~\ref{fig:pulse}(c) shows the pulses at a sufficient time  after  interaction. 
The reflected (transmitted) pulse is seen in the left (right) to the film.
The insets show spectra for the respective pulses. It can be clearly seen that the 
reflected and the transmitted pulses include HHG signals.

\begin{figure*}
    \includegraphics[keepaspectratio,width=\textwidth]{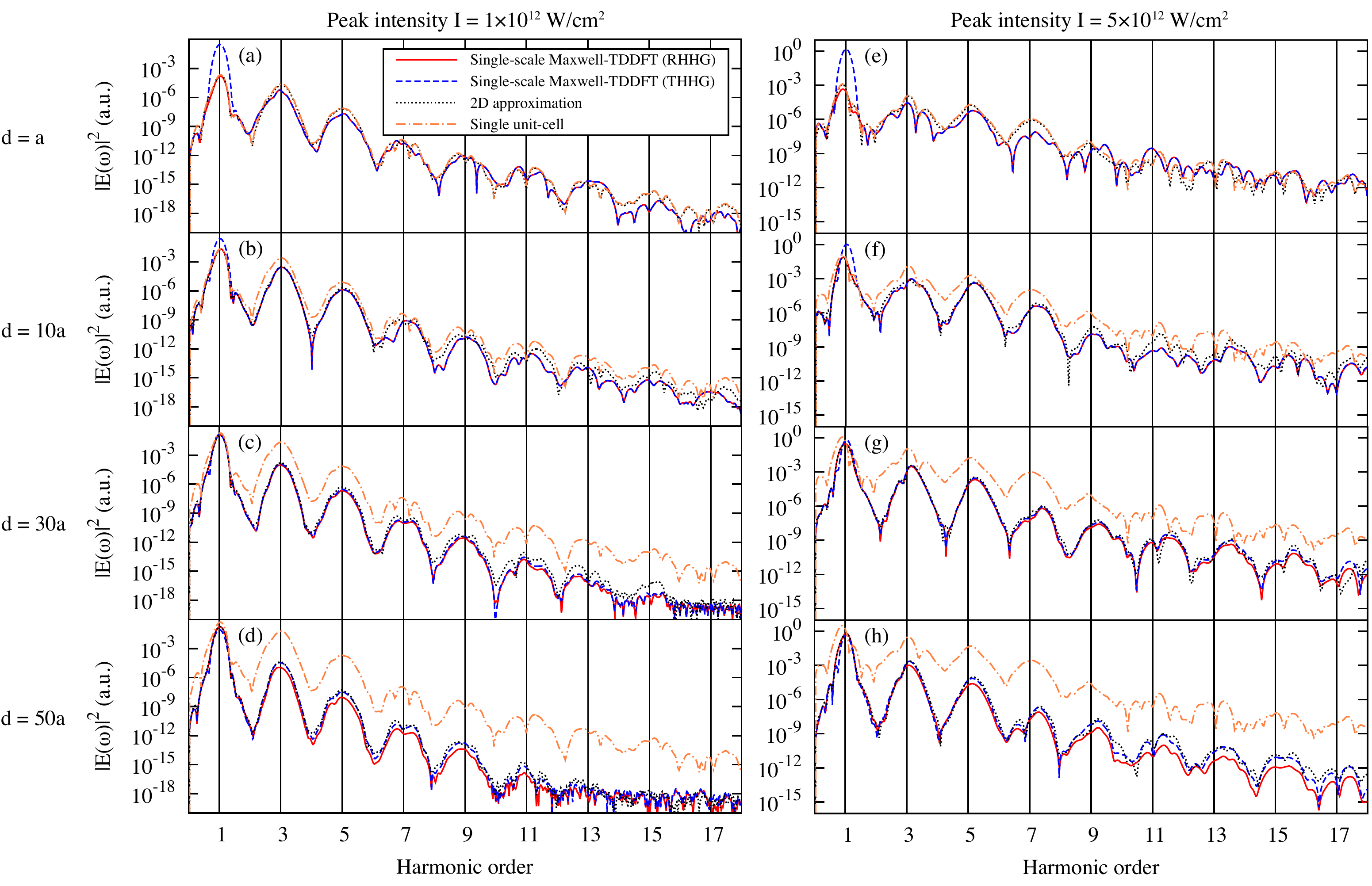}
    \caption{\label{fig:hhg} 
    HHG spectra in the reflected pulse (RHHG) and the transmitted pulse (THHG).
    Left panels [(a)--(d)] show the spectra for a pulse with the peak intensity of $I=1 \times 10^{12}$ W/cm$^2$ and
    right panels [(e)--(h)] for a pulse of $I=5 \times 10^{12}$ W/cm$^2$.
    Top panels [(a), (e)] show the spectra for the film thickness of $d=a$ (0.543 nm), second panels [(b), (f)] for $d=10a$ (5.43 nm), 
    third panels [(c), (g)] for $d=30a$ (16.29 nm), and bottom panels [(d), (h)] for $d=50a$ (27.15 nm).
    The red solid (blue dashed) line for the RHHG (THHG) using the single-scale Maxwell-TDDFT method,
    blue dotted line using the 2D approximation, and orange dash-dotted line using the single unit-cell method.
    }
\end{figure*}

We first discuss thin films with thicknesses less than a few tens of nanometers.
For such films, it is possible to achieve calculations using the single-scale Maxwell-TDDFT method.
Figure \ref{fig:hhg} shows the HHG spectra included in the reflected (RHHG)  and  transmitted (THHG) pulses.
The left panels [Fig.~\ref{fig:hhg}(a)--(d)] show the spectra for the incident pulse with a peak intensity of $I=1\times 10^{12}$ W/cm${}^2$,
and the right panels [Fig.~\ref{fig:hhg}(e)--(h)] show those for a peak intensity of $5\times 10^{12}$ W/cm${}^2$.
Panels (a) and (e) show results for the film of thickness $d=a$ (0.543 nm), (b) and (f) for $10a$ (5.43 nm), 
(c) and (g) for $30a$ (16.29 nm), and (d) and (h) for $50a$ (27.15 nm).
The red solid line (blue dashed line) shows RHHG (THHG) using the single-scale Maxwell-TDDFT method,
and the black dotted line shows HHG using the 2D approximation [Eq.~(\ref{eq:2Dapprox})].
In the 2D approximation, the HHG spectra in the reflected and  transmitted pulses coincide,
except for the component of the fundamental frequency $\omega_0$.
The orange dash-dotted line shows the HHG spectrum using the single unit-cell method.

First, we focus on the extremely thin case of  $d=a$ (0.543 nm) [Fig.~\ref{fig:hhg}(a), (e)]. 
This is a film composed of two atomic layers. Although fabrication of such thin material composed of Si is not easy, 
HHG of such thin materials is highly concerned, as measurements and calculations have been reported for
a number of 2D materials such as graphene and transition metal dichalcogenides \cite{Yoshikawa2017, Yoshikawa2019, Sato2021}.
Here the RHHG and the THHG spectra using the single-scale Maxwell-TDDFT method coincide  accurately, 
as expected from the argument in the 2D approximation [Eq.~(\ref{eq:2Dreftrans})].
However, the HHG spectrum by the 2D approximation (black dotted line) does not coincide accurately with that using 
the single-scale Maxwell-TDDFT method. This indicates the significance of the effect of the surface electronic structure 
that is not included in the 2D approximation.
We note that the validity of the relation of Eq.~(\ref{eq:2Dreftrans}) does not suffer by the surface electronic structure, 
as it can be derived directly from the single-scale Maxwell-TDDFT method\cite{Yamada2018}.
The result of the single unit-cell method (orange dash-dotted line) coincides with the 2D approximation,
indicating that the conductive effect in the film is not important in such extremely thin film.

The above observations hold also at the film of thickness $d=10a$ (5.43 nm) [Fig.~\ref{fig:hhg}(b), (f)], 
except that the discrepancy between spectra using the 2D approximation and the single unit-cell method becomes apparent.
It implies that the effect of the conductivity in the thin film, use of $A^{\rm (t)}$, not the incident pulse $A^{\rm (i)}$, 
to evaluate the current in the right hand side of Eq.~(\ref{eq:2Dapprox}), increases as the film thickness increases, 
because the effect is proportional to $d$.

In the case of $d=30a$ (16.29 nm) [Fig.~\ref{fig:hhg}(c), (g)], the results using the single-scale Maxwell-TDDFT method 
and those using the 2D approximation agree reasonably with each other.
This indicates that neither the surface electronic structure nor the propagation effect is significant at this thickness. 
Only the conductive effect is important.
By closely observing the difference between the RHHG and THHG spectra, the latter is slightly greater than the former.
This indicates that the propagation effect starts to appear.

In the case of $d=50a$ (27.15 nm) [Fig.~\ref{fig:hhg}(d), (h)], it is clear that, in the single-scale Maxwell-TDDFT method, 
the THHG is greater than the RHHG due to the propagation effect.
Now the calculation using the single unit-cell method greatly differs from the others, despite the 2D approximation still reasonably 
reproduces the results using the single-scale Maxwell-TDDFT method.
Looking into detail, the spectrum of the 2D approximation is closer to the THHG in the single-scale Maxwell-TDDFT method than the RHHG.
The high-order part of the spectrum shown in Fig.~\ref{fig:hhg}(d) [and partially Fig.~\ref{fig:hhg}(c)] looks noisy,
presumably because the driving field is too weak ($I=1\times 10^{12}$ W/cm${}^2$) to produce high-order signals at this thickness.

Overall, the 2D approximation is a reasonable approximation to the single-scale Maxwell-TDDFT method in this thickness region.
This indicates that the surface electronic structure and the propagation effects are not significant.
The large difference between the results using the 2D approximation and those using the single unit-cell method indicates that the
conductive effect is significant, even in these very thin films.
 
By comparing the different thicknesses, especially Fig.~\ref{fig:hhg}(e) and (f)--(h), we realize that the HHG spectrum is unclear
in films of thickness $d=a$ (0.543 nm) and becomes clearer as the thickness increases. 
This trend is common in both spectra using the single-scale Maxwell-TDDFT method and the 2D approximation.
It implies that the conductive effect, the current in the thin film that appears in the right hand side of Eq.~(\ref{eq:2Dapprox}), 
contributes to produce a cleaner HHG signal.
This is consistent with the observation in Ref~\onlinecite{Floss2018} where both the
conductive and the propagation effects are included.

\subsection{Thickness dependence}

\begin{figure*}
    \includegraphics[keepaspectratio,width=\textwidth]{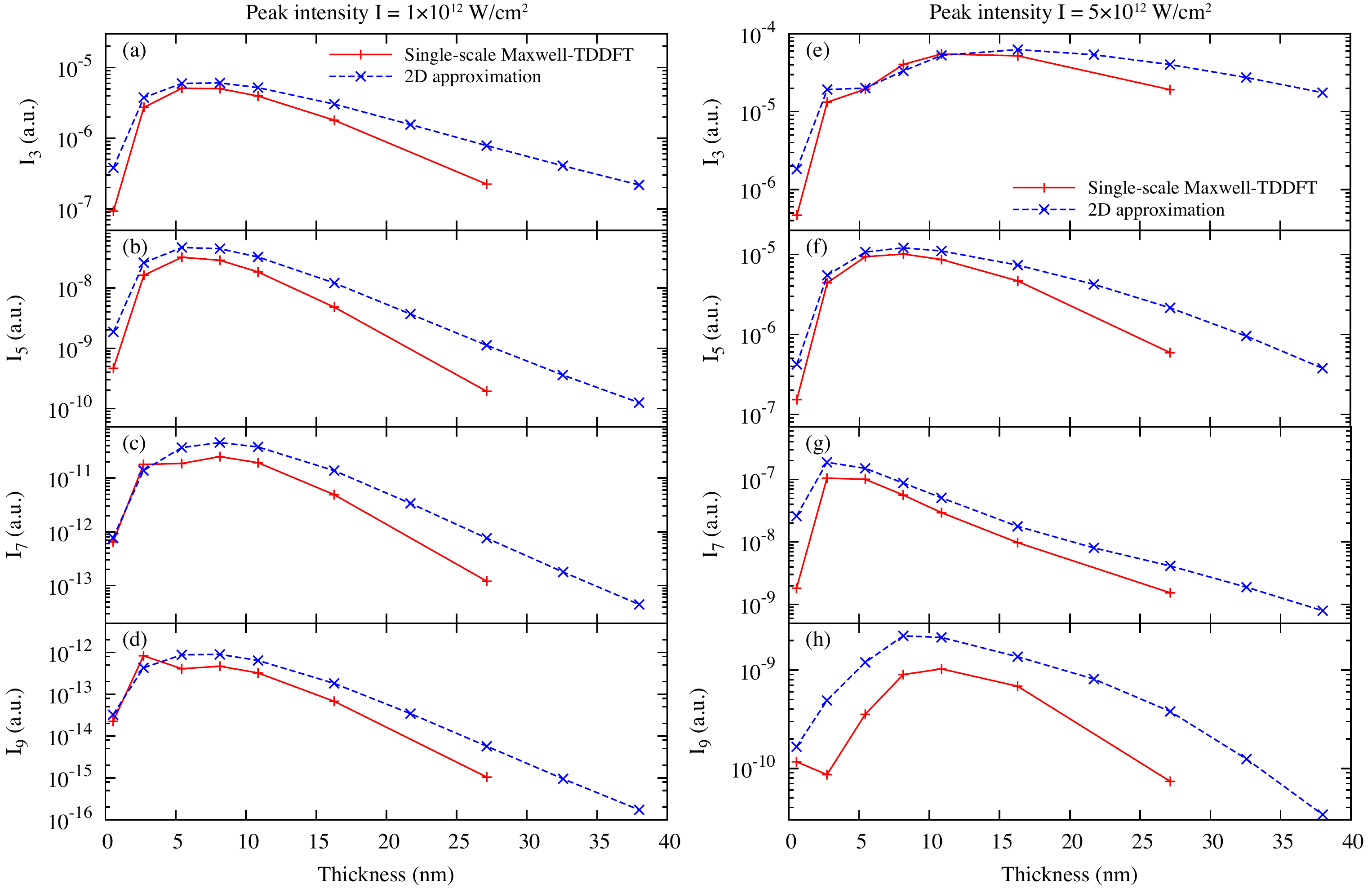}
    \caption{\label{fig:thickness} 
    Thickness dependence of HHG intensities is shown from 3rd [(a), (e)], 5th [(b), (f)], 7th [(c), (g)], and 9th [(d), (h)] orders 
    included in the reflected pulse.
    Left panels [(a)--(d)] for the incident pulse of the peak intensity at $I=1 \times 10^{12}$W/cm$^2$,
    and right panels [(e)--(h)] at $I=5 \times 10^{12}$W/cm$^2$.
    Results using single-scale Maxwell-TDDFT method are shown by red solid lines, and results using
    2D approximation are shown by blue-dashed lines.
    }
\end{figure*}

Figure \ref{fig:thickness} shows the thickness dependence of the intensity of HHG in the reflected pulse using Eq.~(\ref{eq:harmonics}).
The peak intensity of the incident pulses is set at  $I=1\times 10^{12}$ W/cm${}^2$ in Fig.~\ref{fig:thickness}(a)--(d)  and  at $5\times 10^{12}$ W/cm${}^2$ in Fig.~\ref{fig:thickness}(e)--(h).
Intensities of harmonic order of 3rd [(a), (e)], 5th [(b), (f)], 7th [(c), (g)], and 9th [(d), (h)] are shown.
The red solid line (blue dashed line) corresponds to the single-scale Maxwell-TDDFT method (2D approximation).

As seen from Fig.~\ref{fig:thickness}, the thickness dependence of HHG intensities using the 2D approximation (blue dashed line) 
shows qualitative agreement with those using the single-scale Maxwell-TDDFT method (red solid line).
In both calculations, the HHG intensity is maximum around  thicknesses of $d=$ 2--15 nm, irrespective of the order of the HHG.
The appearance of the maximum at these thicknesses can be understood as a consequence of the conductive effect,
as described below, extending the formalism of the 2D approximation.

By taking the Fourier transform of Eq.~(\ref{eq:2Dapprox}), we obtain
\begin{equation}
    -i\omega A^{\rm (t)} (\omega)=  -i\omega A^{\rm (i)} (\omega) + 2\pi d\,\, J[A^{\rm (t)} ](\omega).
\end{equation}
We decompose the current density into linear and nonlinear components as
$ J[A^{\rm (t)}](\omega) = J_{\rm L}[A^{\rm (t)}](\omega)+J_{\rm NL}[A^{\rm (t)}](\omega)$
and use the constitutive relation for the linear part,
\begin{equation}
J_{\rm L}[A^{\rm (t)} ](\omega)=\sigma(\omega) E^{\rm (t)}(\omega)=\frac{i\omega}{c} \sigma(\omega) A^{\rm (t)}(\omega),
\end{equation}
where $\sigma(\omega)$ is the conductivity of the bulk medium.
Combining the above equations, we obtain
\begin{equation}
     \left( 1+\frac{2 \pi d}{c} \sigma(\omega) \right) A^{\rm (t)} (\omega)=   A^{\rm (i)} (\omega) + \frac{2\pi i d}{\omega} J_{\rm NL}[A^{\rm (t)}](\omega).
\end{equation}
We also decompose the transmitted vector potential into linear and nonlinear components \cite{Yamada2018},
$A^{\rm (t)}(\omega) = A_{\rm L}^{\rm (t)}(\omega) + A_{\rm NL}^{\rm (t)}(\omega)$,
where the linear vector potential is given by
\begin{equation}
      A^{\rm (t)}_{\rm L} (\omega) = \left( 1+\frac{2 \pi d}{c} \sigma(\omega) \right)^{-1}  A^{\rm (i)} (\omega).
      \label{eq:Atrans_L}
\end{equation}
The nonlinear term $A^{\rm (t)}_{\rm NL}$ satisfies the following relation:
\begin{equation}
      A^{\rm (t)}_{\rm NL} (\omega)=   \frac{2\pi i d}{\omega}\left( 1+\frac{2 \pi d}{c} \sigma(\omega) \right)^{-1}  J_{\rm NL}[A^{\rm (t)}_{\rm L} +A^{\rm (t)}_{\rm NL} ](\omega).
      \label{eq:Atrans_NL}
\end{equation}
We note that the nonlinear component of the reflected pulse is also given by Eq.~(\ref{eq:Atrans_NL}).
So far, no approximations have been made besides the 2D approximation. 
Here, we assume that the nonlinear component $A^{\rm (t)}_{\rm NL}$ is rather small
and can be ignored to evaluate the nonlinear current density in the right hand side of Eq.~(\ref{eq:Atrans_NL}). 
Then, we have
\begin{equation}
      A^{\rm (t)}_{\rm NL} (\omega)\simeq  \frac{2\pi i d}{\omega}\left( 1+\frac{2 \pi d}{c} \sigma(\omega) \right)^{-1}   J_{\rm NL}[A^{\rm (t)}_{\rm L} ](\omega).
      \label{eq:2Dapprox_lr}
\end{equation}

To investigate how the HHG intensity included in the above $A^{\rm (t)}_{\rm NL}(\omega)$ depends on $d$, we introduce two further assumptions. 
First, we assume that the incident pulse $A^{\rm (i)}(t)$ has a well-defined frequency, which we denote as $\omega_0$. 
Second, the $n$th-order nonlinear current density is proportional to $A^n$,  where $A$ is the amplitude of the vector potential.  
Then, from Eq.~(\ref{eq:Atrans_L}), the $d$ dependence of the $n$th-order term in $J_{\rm NL}[A^{\rm (t)}_{\rm L}](\omega)$ 
is given by $(1+2\pi   \sigma(\omega_0) \,d /c )^{-n}$.
Then, from Eq.~(\ref{eq:2Dapprox_lr}), the $n$th-order term of $A^{\rm (t)}_{\rm NL}(\omega = n\omega_0)$ is proportional to the following factor:
\begin{equation}
A^{\rm (t)}_{\rm NL}(n\omega_0) \propto \frac{d}{\left(1+ \frac{2\pi \sigma(n\omega_0)}{c} d \right) \left(1+ \frac{2\pi \sigma(\omega_0)}{c} d \right)^{n}} .
\label{eq:peak}
\end{equation}
From this $d$ dependence, we expect that the intensity of the HHG of each order shows a maximum as a function of $d$, 
and the peak position is determined by the bulk conductivity.

The conductivity at the frequency of the HHG, $\sigma(n\omega_0)$, depends on the order $n$ and becomes small at high orders, $n>>1$.
For simplicity, we consider the $d$-dependence caused by the conductivity at the fundamental frequency, $\sigma(\omega_0)$.
We expect the peak of HHG appears at the thickness,
\begin{equation}
d \sim \frac{c}{2\pi \sqrt{n-1} \vert \sigma(\omega_0) \vert },
\end{equation}
where we used $\sigma(\omega_0)$ as a pure imaginary number at the frequency $\hbar\omega_0 = 1.5$ eV.
The value $c/(2\pi |\sigma(\omega_0)|) \simeq 19.2$ nm is derived from the conductivity of Si at the frequency $ \omega_0$.
This explains the appearance of the peak around $d=$ 2--15 nm.

\subsection{Propagation effect}

\begin{figure}
    \includegraphics[keepaspectratio,width=\columnwidth]{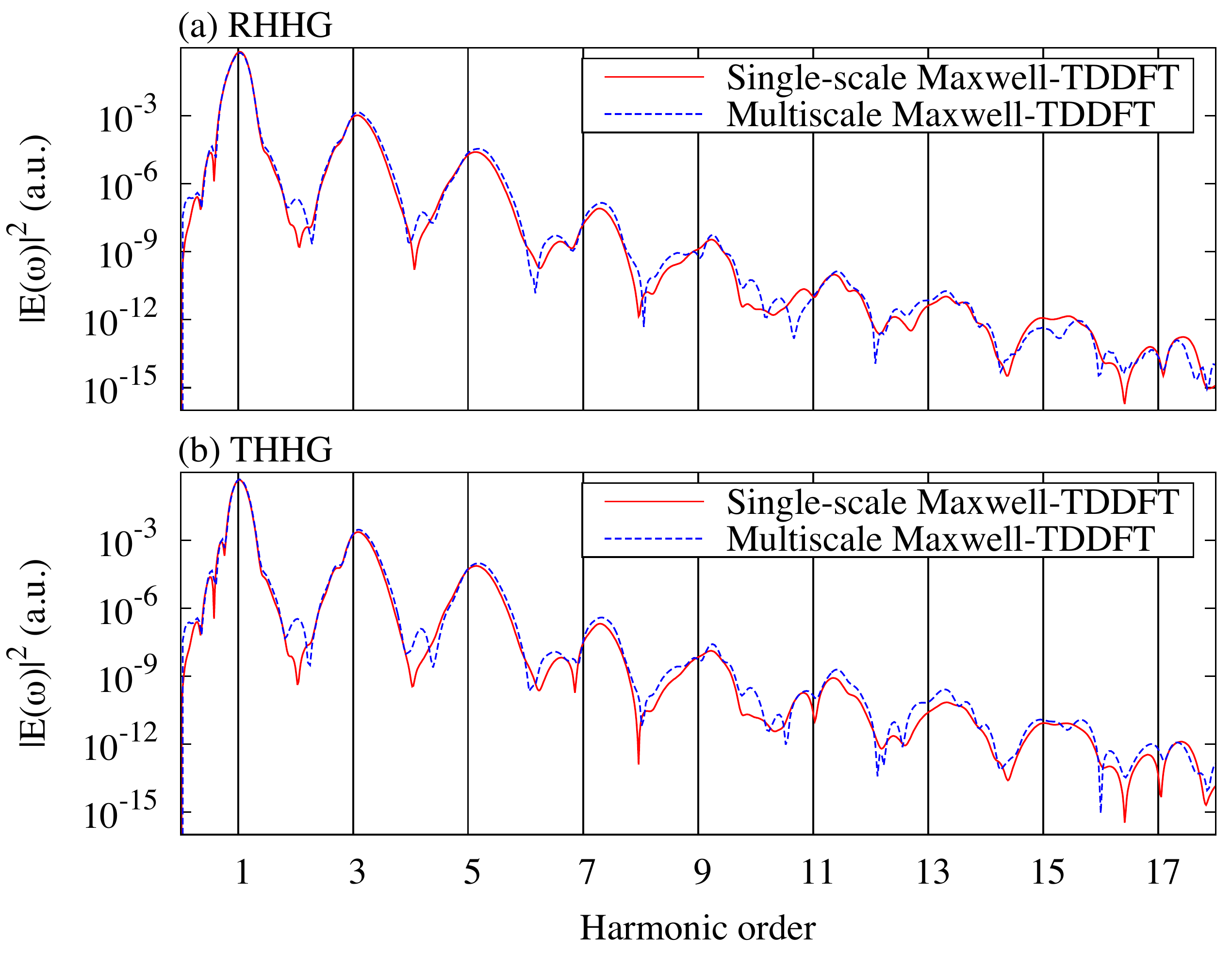}
    \caption{\label{fig:hhg_micro_multi} 
    HHG spectra of $d=50a$ (27.19 nm) film included in the reflected (a) and the transmitted (b) pulses are shown.
    Red solid lines show the  results using the single-scale Maxwell-TDDFT method, whereas the blue dashed lines show those using the multiscale Maxwell-TDDFT method.
    }
\end{figure}

As the film thickness increases, the computational cost of the single-scale Maxwell-TDDFT method rapidly increases.
Therefore, the use of the multiscale Maxwell-TDDFT method is appropriate and necessary.
To confirm the accuracy of the multiscale method, we compare the two methods in Fig.~\ref{fig:hhg_micro_multi}
for RHHG and THHG emitted from a film of thickness $d=50a$ (27.19 nm) and the incident pulse with  maximum intensity  $I=5\times 10^{12}$ W/cm${}^2$,
which is the same conditions as those in Fig.~\ref{fig:hhg}(h).
Because the spectra using two methods coincide accurately with each other, we may conclude that the multiscale Maxwell-TDDFT method,
which ignores the effect of the surface electronic structure, is reliable for films of this thickness and greater.

We note that, as mentioned in Sec.~II.C, the multiscale Maxwell-TDDFT method is identical to the 2D approximation for
very thin films where the macroscopic electric field may be regarded as uniform inside the thin film.
As discussed in Fig.~\ref{fig:hhg}, the 2D approximation was a good approximation for films of thickness less than $d=50a$ (27.19 nm).
In summary, we expect  the multiscale Maxwell-TDDFT method to be reliable for thin films of any thickness as long as
the surface electronic structure is not important.

\begin{figure}
    \includegraphics[keepaspectratio,width=\columnwidth]{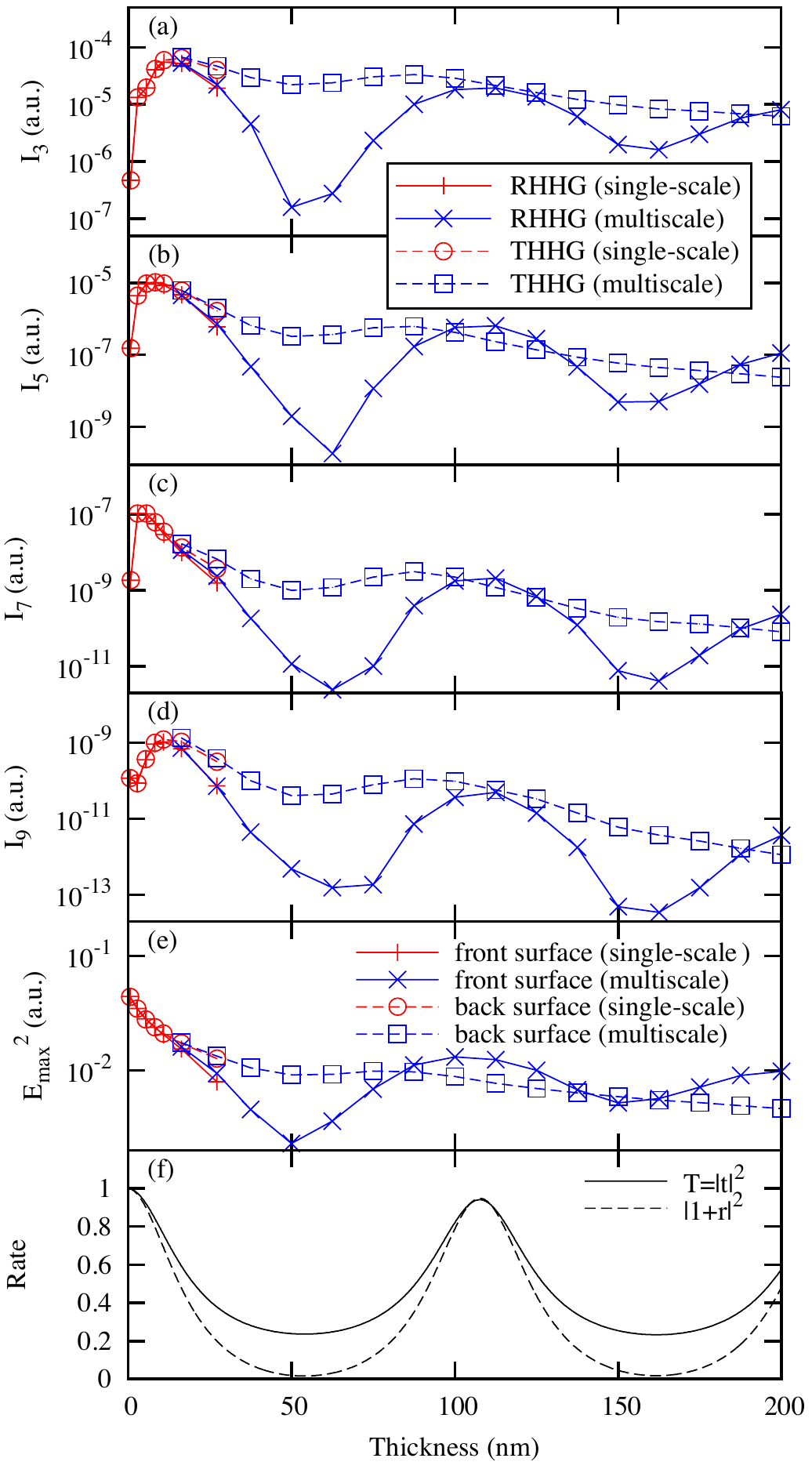}
    \caption{\label{fig:thickness_multi} 
    (a)--(d) Thickness dependence of the intensities of the reflected (solid lines with crosses) and the transmitted (dashed lines with squares)
    HHG, from 3rd (a) to 9th (d) order. Red lines and symbols are calculated using the single-scale Maxwell-TDDFT
    method, and blue lines and symbols using the multiscale Maxwell-TDDFT method.
    (e) Maximum intensities of the electric fields averaged over the 2D area of the front (back) surface are shown by red lines with symbols (blue lines with symbols).
    (f) Intensities of electric fields in ordinary electromagnetism at the front (back) surface are shown by  dashed (solid) lines.
    }
\end{figure}

Figure~\ref{fig:thickness_multi}(a)--(d) shows the thickness dependence of the third- to ninth-order harmonics
included in the reflected and transmitted pulses for the incident pulse with a peak intensity of $I=5 \times 10^{12}$ W/cm$^2$.
Results using the single-scale Maxwell-TDDFT method are shown for  thickness $d \leq 50a$ (27.19 nm). 
The RHHG shown here are the same as those in Fig.~\ref{fig:thickness}(e)--(h).
The results using the multiscale Maxwell-TDDFT method are shown for  thickness $d \geq 30a$ (16.29 nm).

Figure~\ref{fig:thickness_multi}(e) shows the square of the maximum electric field amplitude in time, $E_{\rm max}=\max_t |E(t)|$,  
at the front surface and the back surface of the film.
When using the single-scale Maxwell-TDDFT method, an average over the surface is taken.
Figure~\ref{fig:thickness_multi}(f) shows $|t|^2$ and $|1+r|^2$, where $t$ and $r$ are the amplitude of 
transmission and reflection coefficients in ordinary electromagnetism:
\begin{equation}
r = \frac{r_0(1-e^{2i\phi})}{1-r_0^2 e^{2i\phi}},
\hspace{5mm}
t = \frac{ (1-r_0^2) e^{i\phi}}{1-r_0^2 e^{2i\phi}},
\end{equation}
where $r_0=(1-n)/(1+n)$ and $\phi=2\pi nd/\lambda$.
The index of refraction $n$ of Si and the wavelength $\lambda$ is evaluated at the frequency $\omega_0$.
$T=|t|^2$ is equal to the transmittance and is equal to the square of the electric field at the back surface,
while $|1+r|^2$ is equal to the square of the electric field at the front surface.

Because the frequency $\omega_0$ is below the direct bandgap of Si, $|t|^2$ and $|1+r|^2$
become unity when $2\pi nd$ is equal to the wavelength and its integer multiples.
As seen from Fig.~\ref{fig:thickness_multi}(e), however, the magnitude of the electric field at the back surface 
decreases monotonically as the film thickness increases in the Maxwell-TDDFT calculations.
The magnitude of the electric field at the front surface shows more oscillatory behavior, with a clear
minimum at $\pi d = \lambda$. However, the magnitude is smaller at $\pi d = 2\lambda$ than at the zero-thickness limit.
These decreases originate the nonlinear propagation effect included in both the single-scale and multiscale 
Maxwell-TDDFT methods.
In any case, the magnitude of the electric field at neither the front nor the back surface exceeds the
magnitude of the electric field at the zero-thickness limit.

The intensities of RHHG and THHG coincide with each other when the film is sufficiently thin.
As  discussed previously, they are well understood by Eq.~(\ref{eq:2Dreftrans}) and start to deviate at a distance of 10--20 nm.
As seen from Fig.~\ref{fig:thickness_multi}, except the $d < 20$ nm region, the thickness dependence of RHHG (THHG) is correlated 
faithfully with the maximum electric field at the front (back) surface. 
This indicates that the magnitude of the RHHG and the THHG is  determined simply by the intensity of its 
driving electric field at the respective surfaces.
Because the magnitude of the electric field at the front or  back surface does not exceed those
at the zero-thickness limit, the maximum of the intensity of the HHG signals also appears around $d \sim 10$ nm
as seen in Fig.~\ref{fig:thickness}.


\section{Conclusion \label{sec:conclusion}}

We have developed a few theoretical and computational methods based on first-principles
TDDFT and investigated HHG from thin films of crystalline solids with thicknesses from 
a few atomic layers to a few hundreds of nanometers. 
Using these methods, it is possible to investigate effects of
(1) surface electronic structure, (2) conductivity of the film, and (3) light propagation
of fundamental and high-harmonic fields, as well as (4) the effect of electronic motion
in the bulk energy band.

Among methods used herein, the most sophisticated method is the single-scale
Maxwell-TDDFT method, which considers all four effects mentioned above.
It can be applied to thin films with thicknesses less than a few tens of nanometers.
For thicker films, the multiscale Maxwell-TDDFT method, in which a coarse-graining 
approximation is introduced, is applicable. It can treat the effects except for the surface
electronic structure. To achieve insight into the thickness dependence of the HHG, 
 the 2D approximation is useful. We applied these methods to thin films of crystalline silicon 
as a prototype material, and obtained the following conclusions.

For thin films with thicknesses less than a few tens of nanometers, HHG spectra of reflected and
transmitted pulses are almost identical. This could be understood from the 2D approximation.
In extremely thin films, the HHG intensity increases as the thickness of the film increases.
However, the intensity soon saturates and reaches its maximum at a thickness of around
2--15 nm. The saturation of the intensity originates from the current that flows in the thin film
and can be described using the bulk linear conductivity of the medium.
This conductive effect also creates a clean HHG spectrum.

As the thickness increases beyond a few tens of nanometers, it is found that the
intensity of the reflected (transmitted) HHG  strongly correlates with the intensity
of the electric field at the front (back) surface of the thin film.
By reflecting the interference effect in  pulse propagation,
we find an enhancement in the reflected HHG when the film thickness is equal to 
integer or half-integer multiples of the fundamental wavelength in the medium, 
$\lambda/n$, where $n$ is the index of refraction.
However, the transmitted HHG monotonically decreases as the thickness
increases, owing to the nonlinear propagation of the fundamental wave.

\begin{acknowledgements}
This research was supported by JST-CREST under grant number JP-MJCR16N5, and
by MEXT Quantum Leap Flagship Program (MEXT Q-LEAP) Grant Number
JPMXS0118068681, and by JSPS KAKENHI Grant Number 20H2649.
Calculations are carried out at Oakforest-PACS at JCAHPC under the support 
through the HPCI System Research Project (Project ID: hp20034), and
Multidisciplinary Cooperative Research Program in CCS, University of Tsukuba.
\end{acknowledgements}

\appendix

%


\end{document}